\documentclass[aps, prl, superscriptaddress, floatfix, longbibliography, twocolumn]{revtex4-2}

\usepackage{amsfonts, amssymb, amsmath}
\usepackage{graphicx}
\usepackage{float}
\usepackage[plain]{fancyref}
\usepackage{placeins}
\usepackage{hyperref}
\usepackage{color,soul}
\usepackage{siunitx}
\usepackage{natbib}
\usepackage[version=3]{mhchem}
\usepackage{physics}
\usepackage{pdfrender}
\usepackage{mathtools}
\usepackage{cases}
\usepackage{verbatim}
\definecolor{myorange}{cmyk}{0.0149, 0.7523, 0.7816, 0.003}
\definecolor{mypurple}{cmyk}{0.6638, 0.9652, 0, 0}

\DeclareSIUnit\angstrom{\text {Å}}
\newcommand{\VQDi}{\ensuremath{V_\mathrm{QD1}}}
\newcommand{\VQDii}{\ensuremath{V_\mathrm{QD2}}}
\newcommand{\VQDiii}{\ensuremath{V_\mathrm{QD3}}}
\newcommand{\VL}{\ensuremath{V_\mathrm{L}}}
\newcommand{\VR}{\ensuremath{V_\mathrm{R}}}
\newcommand{\VSi}{\ensuremath{V_\mathrm{S1}}}
\newcommand{\VSii}{\ensuremath{V_\mathrm{S2}}}
\newcommand{\Vb}{\ensuremath{V_\mathrm{b}}}
\newcommand{\IL}{\ensuremath{I_\mathrm{L}}}
\newcommand{\IR}{\ensuremath{I_\mathrm{R}}}
\newcommand{\ISi}{\ensuremath{I_\mathrm{S1}}}
\newcommand{\ISii}{\ensuremath{I_\mathrm{S2}}}
\newcommand{\QDi}{\ensuremath{\mathrm{QD}_1}}
\newcommand{\QDii}{\ensuremath{\mathrm{QD}_2}}
\newcommand{\QDiii}{\ensuremath{\mathrm{QD}_3}}
\newcommand{\NL}{\ensuremath{\mathrm{N}_\textrm{L}}}
\newcommand{\NR}{\ensuremath{\mathrm{N}_\textrm{R}}}
\newcommand{\SL}{\ensuremath{\mathrm{S}_\textrm{1}}}
\newcommand{\SR}{\ensuremath{\mathrm{S}_\textrm{2}}}

\begin{document}

\title{Crossed Andreev reflection and elastic co-tunneling in a three-site Kitaev chain nanowire device}

\author{Alberto~Bordin}
\author{Xiang~Li}
\author{David~van~Driel}
\author{Jan~Cornelis~Wolff}
\author{Qingzhen~Wang}
\author{Sebastiaan~L.~D.~ten~Haaf}
\author{Guanzhong~Wang}
\author{Nick~van~Loo}
\author{Leo~P.~Kouwenhoven}
\email{l.p.kouwenhoven@tudelft.nl}
\author{Tom~Dvir}
\email{tom.dvir@gmail.com}
\affiliation{QuTech and Kavli Institute of NanoScience, Delft University of Technology, 2600 GA Delft, The Netherlands}

\date{\today}

\begin{abstract}
The formation of a topological superconducting phase in a quantum-dot-based Kitaev chain requires nearest neighbor crossed Andreev reflection and elastic co-tunneling. Here we report on a hybrid InSb nanowire in a three-site Kitaev chain geometry --- the smallest system with well-defined bulk and edge --- where two superconductor-semiconductor hybrids separate three quantum dots. We demonstrate pairwise crossed Andreev reflection and elastic co-tunneling between both pairs of neighboring dots and show sequential tunneling processes involving all three quantum dots. These results are the next step towards the realization of topological superconductivity in long Kitaev chain devices with many coupled quantum dots.
\end{abstract}

\maketitle
\newpage

\section*{Introduction}

The Kitaev chain was proposed over two decades ago as a platform that supports unique non-local excitations known as Majorana bound states \cite{kitaev2001unpaired}. Proposals~\cite{Sau.2012, Leijnse.2012, Fulga.2013} for the realization of such a Kitaev chain rely on creating an array of spin-polarized quantum dots (QDs) where neighboring QDs are coupled via two mechanisms: elastic co-tunneling (ECT) and crossed Andreev reflection (CAR). ECT involves the hopping of a single electron between two QDs. In the process of CAR, two electrons from neighboring QDs simultaneously enter a superconductor to form a Cooper pair, or, in reversed order, two electrons forming a Cooper pair are split into two QDs \cite{Recher.2001,lesovik2001electronic,sauret2004quantum}. Experiments have so far focused on chains consisting of two QDs, showing both CAR and ECT in such systems \cite{Hofstetter.2009jdl,Herrmann2010carbon,wei2010positive,hofstetter2011finite,das2012high,schindele2012near,Schindele.2014,tan2015, Gramich.2017,baba2018cooper,scherubl2020large,ranni2021real,Wang.2022,Wang.2022.2DEG} and even strong coupling the QDs to form a minimal Kitaev chain \cite{dvir2022realization}. Longer QD chains, necessary for the formation of a topological phase, have so far not been realized and are the long-term goal of this research.

In this work, we report on the fabrication of a three-site Kitaev device and its characterization at zero magnetic field. We show CAR and ECT between each pair of neighboring QDs and show that transport across the entire device is possible through sequential events of CAR and ECT. By measuring the currents on all of the terminals of our device, we can identify all the possible CAR and ECT combinations.

\section*{Device structure}
\begin{figure}[ht!]
    \centering
    \includegraphics[width=\columnwidth]{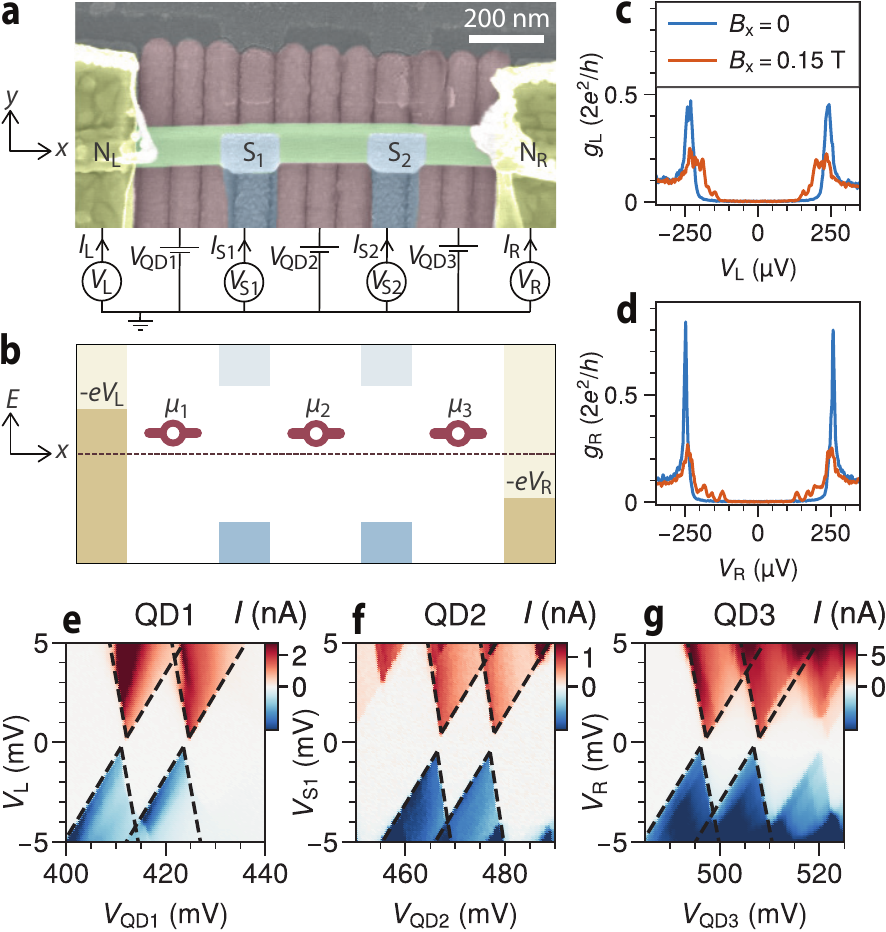}
    \caption{\textbf{a.} False-colored scanning electron micrograph of the measured device. An InSb nanowire (green) is deposited on top of 11 finger gates (red) and it is contacted with two superconducting leads S$_1$ and S$_2$ (blue) and two normal leads $\NL$ and $\NR$  (yellow). Every contact is connected to a corresponding voltage bias source and current meter.
    \textbf{b.} Illustrative energy diagram. Brown symbols represent QD energy levels when occupied by an electron.
    \textbf{c., d.} Spectroscopy of the hybrid segments. $g_\mathrm{L} \equiv \frac{dI_\mathrm{L}}{dV_\mathrm{L}}$ and $g_\mathrm{R} \equiv \frac{dI_\mathrm{R}}{dV_\mathrm{R}}$ are obtained by numerical differentiation of the DC currents. Gate settings are reported in Supplementary Information (Fig.~S1,~S2~and~S3).
    \textbf{e., f., g.} Coulomb diamond characterization of $\QDi$ (panel e), $\QDii$ (panel f), $\QDiii$ (panel g). Fitting to a constant interaction model \cite{Hanson.2004} yields charging energies of 4, 3.5, $\SI{3.3}{mV}$ and lever arms of 0.32, 0.33, 0.31 for $\QDi$, $\QDii$, and $\QDiii$ respectively. 
    }
    \label{fig:1}
\end{figure}

In Fig.~\ref{fig:1}a we show a scanning electron micrograph of device A. This device consists of an InSb nanowire placed on top of an array of 11 finger gates separated by a thin dielectric.  Two superconducting Al contacts (marked $\textrm{S}_\textrm{1}$ and $\textrm{S}_\textrm{2}$) are evaporated on top of the nanowire using the shadow-wall lithography technique \cite{Heedt.2021, Mazur.2022}. Both sides of the device are further contacted by two normal Cr/Au probes (marked $\NL$ and $\NR$).  Every contact is connected to an independent voltage source ($\VL$, $\VSi$, $\VSii$, $\VR$) and current meter ($\IL$, $\ISi$, $\ISii$, $\IR$). The two finger gates underneath the semiconductor-superconductor hybrid segments control their chemical potential, while the other 9 gates form QDs on each of the three bare InSb sections. The QD chemical potentials $\mu_1$, $\mu_2$ and $\mu_3$, are controlled by the gate voltages $\VQDi$, $\VQDii$ and $\VQDiii$ respectively (Fig.~\ref{fig:1}b). See Supplementary Information for further nanofabrication details and gate settings.

\section*{Results}
\subsection{Device characterization}
Discrete Andreev bound states (ABS) in a hybrid semiconductor-superconductor segment, separating two QDs, can efficiently mediate CAR and ECT between them \cite{fulop2015magnetic,liu2022tunable,Bordin.2022}. We find such discrete states by controlling the two finger gates underneath contacts S$_1$ and S$_2$. Fig.~\ref{fig:1}c,d show the spectra measured on the first and the second hybrid respectively, by using the finger gates separating the superconducting and normal contacts as tunneling barriers \cite{vanDriel2022spin}. In the absence of an external magnetic field, both hybrid segments show a hard superconducting gap, i.e., lack of subgap transport. A closer inspection of the gate dependence (see Fig.~S2) shows that ABS are present at energies close to the energy gap. At 150 mT magnetic field, the ABS are more visible in the spectrum. The remainder of the experiment was conducted at zero applied magnetic field and fixed value of the hybrid gates.

In Fig.~\ref{fig:1}e-f we characterize QDs 1-3 respectively. The observed Coulomb diamond structure allows us to estimate the charging energy of all QDs to be between 3 and \SI{4}{mV} and the lever arm of the underlying gates to be $\approx 0.3$ (see also Fig.~S3). We also note the presence of a superconducting gap in the spectrum.

\subsection*{Pairwise CAR and ECT between neighboring QDs}
\begin{figure}[ht!]
    \centering
    \includegraphics[width=\columnwidth]{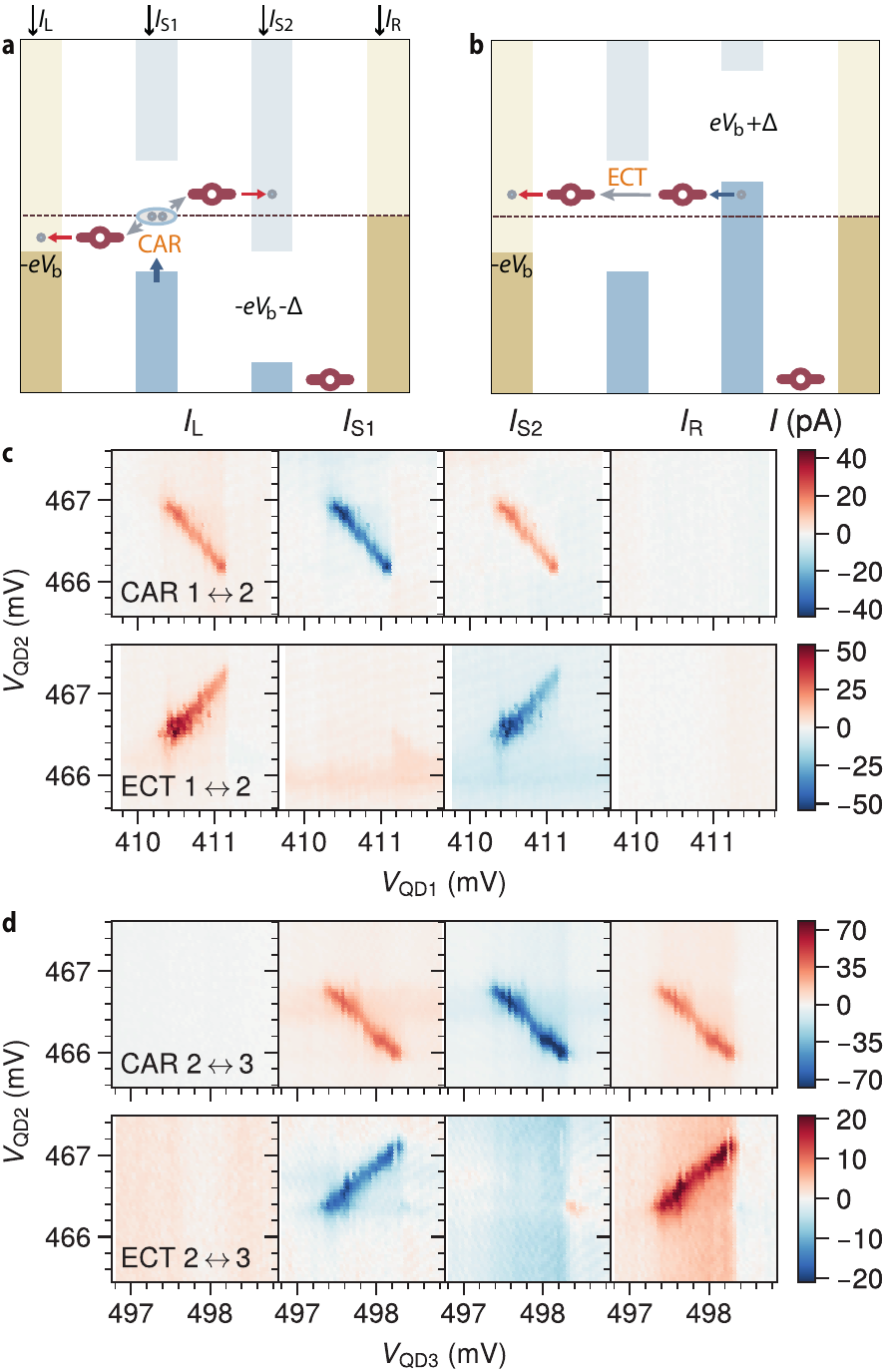}
    \caption{\textbf{a., b.} Schematic diagrams of CAR and ECT processes between $\QDi$ and $\QDii$. CAR is measured by applying $\Vb$ on $\NL$ and $\Vb + \Delta/e$ on $\SR$ (panel a). ECT is measured by applying $\Vb$ on $\NL$ and $-\Vb - \Delta/e$ on $\SR$ (panel b). \textbf{c.} CAR and ECT between $\QDi$ and $\QDii$. The currents $\IL$, $\ISi$, $\ISii$, and $\IR$ are measured as a function of $\VQDi$ and $\VQDii$. $\VL = \Vb = \SI{150}{\micro V}$, while $\VSii =  \Vb+\Delta/e = \SI{380}{\micro V}$ (top row) or $\VSii = -\SI{380}{\micro V}$ (bottom row). \textbf{d.} CAR and ECT between $\QDii$ and $\QDiii$. The currents through the leads as a function of $\VQDii$ and $\VQDiii$ are measured with $\VSi =  \SI{380}{\micro V}$ in the top row and $\VSi =  -\SI{380}{\micro V}$ in the bottom row, while $\VR = \SI{150}{\micro V}$.
    }
    \label{fig:2}
\end{figure}

We begin by demonstrating CAR and ECT processes between pairs of neighboring QDs. 
Fig.~\ref{fig:2}a shows schematically how CAR between $\QDi$ and $\QDii$ is measured while $\QDiii$ is kept off-resonance and thus does not participate in the transport. CAR involves current flowing from a superconductor into the neighboring leads (or vice-versa). In recent works, CAR was measured setting symmetric voltage biases, $\Vb$, on two normal leads on both sides of the hybrid segment~\cite{Wang.2022, Bordin.2022, Wang.2022.2DEG}. 
Here, to account for the presence of the superconducting gap in S$_2$ ($\Delta\approx \SI{230}{\micro eV}$), we apply a bias of $\Vb + \Delta/e$ to the superconducting leads.
In this configuration, CAR can be sustained as long as $\mu_1 = -\mu_2$ and the two chemical potentials are in the bias window $-\abs{e\Vb} < \mu_1,\mu_2 < \abs{e\Vb}$~\cite{Wang.2022}.
ECT can be measured in an anti-symmetric bias configuration. Due to the presence of the superconducting gap, such configuration similarly requires adding $\Delta$ to the bias on S$_2$, as shown schematically in Fig.~\ref{fig:2}b. 

Fig.~\ref{fig:2}c shows the currents through all leads measured in the bias configuration that allows for CAR (top row) and ECT (bottom row) as a function of $\VQDi$ and $\VQDii$. In the top row, we find that the currents $\IL$, $\ISi$, and $\ISii$ are largest along a diagonal line consistent with $\mu_1 = -\mu_2$. Moreover, we note that $\IL$ and $\ISii$ are positive and nearly equal to each other and drain to the ground only through the lead S$_1$. These observations signal the presence of CAR between $\QDi$ and $\QDii$. 
The bottom row is measured in a bias configuration that supports ECT. The measurements show finite $\IL$ and $\ISii$ currents with maxima along a diagonal compatible with $\mu_1 = \mu_2$. In this case, $\IL$ and $\ISii$ have opposite signs, and almost no current flows through S$_1$, indicating ECT processes between $\QDi$ and $\QDii$. 

Analogously, we measure CAR (and ECT) signatures between $\QDii$ and $\QDiii$ by applying effectively symmetric (and antisymmetric) biases $\VSi$ and $\VR$. Measured currents on all leads as a function of $\VQDii$ and $\VQDiii$ are shown in Fig.~\ref{fig:2}d attesting the presence of CAR and ECT between $\QDii$ and $\QDiii$.
We also notice finite currents $< \SI{10}{pA}$ that depend only on the value of $\VQDiii$. We interpret this as a sign of local Andreev reflection (LAR) not being completely suppressed by the charging energy of $\QDiii$ (see Fig.~S3).

The results shown in Fig.~\ref{fig:2} demonstrate both CAR and ECT --- the crucial ingredients of a Kitaev chain --- between every pair of QDs.

To separately detect CAR and ECT between pairs of dots, we exploit above the freedom to apply independent voltage biases to each superconductor. This freedom might not always be accessible, e.g., in a Kitaev chain design with the superconductors connected in a loop. In the following, we discuss the signatures of CAR and ECT when both superconducting leads are grounded. 

\subsection{Two-terminal CAR and ECT processes}

\begin{figure}[ht!]
    \centering
    \includegraphics[width=\columnwidth]{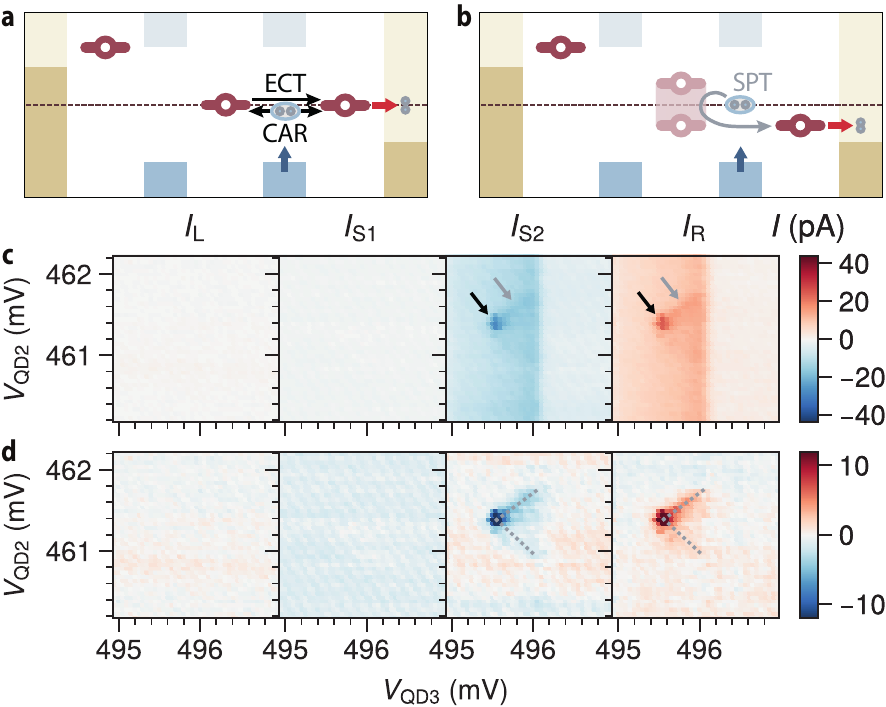}
    \caption{\textbf{a.} Schematic illustration of the resonant CAR and ECT tunneling. When $\mu_2 = \mu_3 = 0$ both CAR and ECT are allowed between $\QDii$ and $\QDiii$, allowing a complete transport cycle to transfer a Cooper pair between $\NR$ and $\SR$. \textbf{b.} Schematic illustration of the Shiba-assisted local pair tunneling (SPT). 
    \textbf{c.} Current through the device as a function of $\VQDii$ and $\VQDiii$, with $\VQDi = \SI{414.1}{mV}$, equivalent to $\mu_1 \approx \SI{230}{\micro eV}$. \textbf{d.} Same data of panel c but with subtracted LAR and saturated colorscale. LAR is extracted from the average of the top and bottom linecuts at fixed $\VQDii$, see the code in the linked repository for more details.}
    \label{fig:3}
\end{figure}

We set $\VSi = \VSii = 0$, and begin by discussing CAR and ECT processes between $\QDii$ and $\QDiii$ while keeping $\QDi$ off-resonance (see schematics in Fig.~\ref{fig:3}a,b). We observe three transport mechanisms involving only leads $\SR$ and $\NR$. 

The first transport mechanism, already mentioned above, is LAR, giving rise to a signal that depends only on the chemical potential of $\QDiii$. 

The second transport mechanism takes place when $\mu_2 = \mu_3 = e\VSii = 0$, as depicted in Fig.~\ref{fig:3}a. In this alignment of the chemical potentials, both CAR and ECT are allowed. A Cooper pair can be transmitted between S$_2$ and $\NR$ by sequential CAR and ECT processes: first, a Cooper pair is split from $\SR$ to both QDs. Then the electron in $\QDiii$ is drained to $\NR$, allowing ECT to shuttle the other electron between $\QDii$ and $\QDiii$, which is finally drained as well. Because of the resonant condition on the chemical potentials, this process appears as a single spot in the measurements shown in Fig.~\ref{fig:3}c (marked by the black arrow).

When $\mu_2 \neq 0$, the resonant CAR-ECT process is not allowed anymore; however, a third transport mechanism might happen. 
A small but finite LAR in $\QDii$ leads to the formation of a Yu-Shiba-Rusinov state~\cite{Yu.1965,shiba1968classical,rusinov1969superconductivity,Bauer.2007,meng.2009,Lee.2014}, which has an electron and a hole part. Both parts can assist either ECT followed by ECT or CAR followed by CAR to transfer a Cooper pair from $\SR$ to $\NR$. This process was previously named ``Shiba-assisted local pair tunneling'' \cite{scherubl2020large,scherubl2022cooper} and is depicted schematically in Fig.~\ref{fig:3}b. 
The Shiba-assisted local pair tunneling occurs when $\QDiii$ is resonant with either the electron or the hole part of the Yu-Shiba-Rusinov state on $\QDii$. This process appears as two current features observing $|\mu_2| = |\mu_3|$, one with a positive slope and the other with a negative slope in the measurement (marked with a grey arrow). 
To better visualize it, we highlight it in Fig.~\ref{fig:3}d with a grey dashed line after subtracting the LAR background.

\subsection*{Three-dot sequential CAR and ECT}

\begin{figure}[ht!]
    \centering
    \includegraphics[width=\columnwidth]{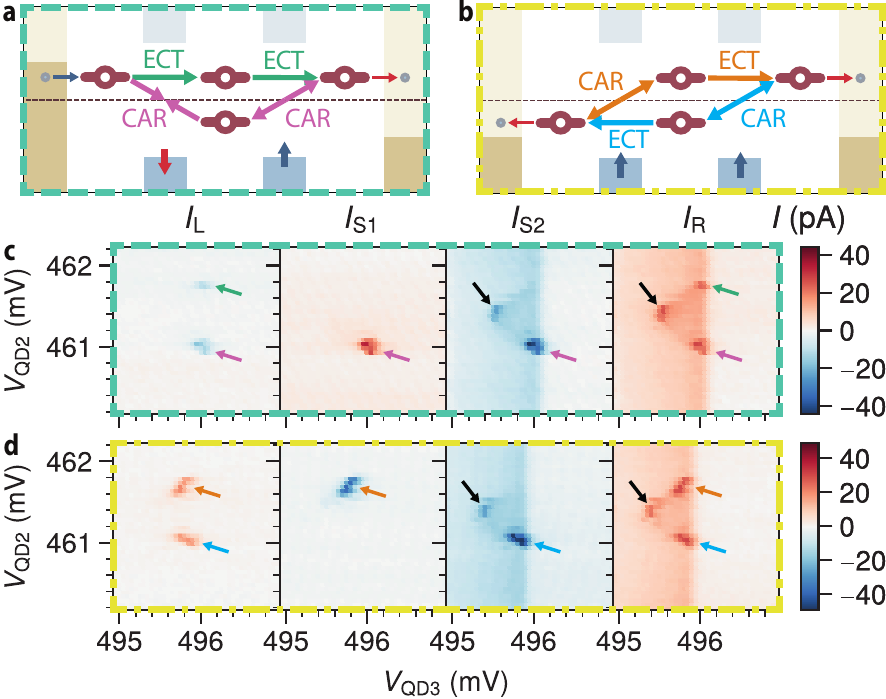}
    \caption{\textbf{a.} Schematic illustration of sequential ECT processes (with $\mu_1 = \mu_2 = \mu_3$, green) and sequential CAR processes (with $\mu_1 = -\mu_2 = \mu_3$, pink). \textbf{b.} Schematic illustration of CAR followed by ECT (with $-\mu_1 = \mu_2 = \mu_3$, orange) and ECT followed by CAR (with $\mu_1 = \mu_2 = -\mu_3$, blue). \textbf{c.} Current through the device as a function of $\VQDii$ and $\VQDiii$, with $\VR=-\VL = \SI{150}{\micro V}$ and $\VQDi = \SI{413.8}{mV}$, equivalent to $\mu_1 \approx \SI{130}{\micro eV}$. \textbf{d.} Current through the device as a function of $\VQDii$ and $\VQDiii$, with $\VR=\VL = \SI{150}{\micro V}$ and $\VQDi = \SI{413}{mV}$, equivalent to $\mu_1 \approx \SI{-100}{\micro eV}$. Note that the arrow colors in panels c and d correspond to the process colors in panels a and b, while the black arrow corresponds to the resonant CAR and ECT process shown in Fig.3a,c.}
    \label{fig:4}
\end{figure}

Setting $\abs{\mu_1} < \abs{e\VL}$ allows $\QDi$ to participate in transport, opening the route for sequential CAR and ECT processes involving all three QDs and four leads. Fig.~\ref{fig:4}a shows schematically such processes with antisymmetric bias settings ($\VL = -\VR$). In this configuration, electrons incoming from $\NL$ can be transferred all the way to $\NR$. This can happen in two ways: either with two sequential ECT processes (marked by green arrows) or with two sequential CAR ones (marked by pink arrows). Sequential ECT events can first transfer an electron from $\QDi$ to $\QDii$ and then from $\QDii$ to $\QDiii$, provided that the QD chemical potentials are all aligned ($\mu_1 = \mu_2 = \mu_3$). Alternatively, if the QD chemical potentials are anti-aligned ($\mu_1 = -\mu_2 = \mu_3$), sequential CAR can begin by transmitting a Cooper pair into $\SL$ and then splitting a Cooper pair from $\SR$, resulting in a net transfer of one electron from $\QDi$ to $\QDiii$. Equivalently, this sequential CAR process can be seen as an electron from $\QDi$ being converted into a hole in $\QDii$ and converted back to an electron into $\QDiii$.

Fig.~\ref{fig:4}c shows the measured currents through our device as a function of $\VQDii$ and $\VQDiii$ for fixed $\VR = -\VL$. These measurements show both the two-QD processes discussed above (see black arrows) and the three-QD processes discussed here. The sequential ECT processes appear as a single spot in $\IL$ and $\IR$, where $\mu_2$ and $\mu_3$ are aligned with $\mu_1$ (marked by the green arrow). Since the superconducting leads drain no current in ECT, $\ISi$ and $\ISii$ do not show this sequence.  The sequence involving two CAR processes (marked by the pink arrow) appears as a spot in the currents measured on all leads when $\mu_3 = \mu_1$ and $\mu_2 = -\mu_1$. The currents alter in sign at every lead, going from negative $\IL$, to positive at $\ISi$, to negative at $\ISii$, and back to positive at $\IR$, corresponding to Cooper pair formation in $\SL$ followed by Cooper pair splitting in $\SR$. 

Under symmetric bias conditions, the current is sustained when both leads $\NL$ and $\NR$ drain electrons (see Fig.~\ref{fig:4}b). The two sequences involving all QDs in agreement with this condition are CAR followed by ECT and the opposite, ECT followed by CAR (Fig.~\ref{fig:4}b). The first sequence of CAR followed by ECT can be seen in Fig.~\ref{fig:4}d in the features marked by the orange arrow. It is seen as a current feature appearing when $\mu_3 \approx \mu_2 \approx -\mu_1$ in $\IL$ and $\IR$. This feature further appears in $\ISi$ but not in $\ISii$, since CAR between $\QDi$ and $\QDii$ drains current to the ground through S$_1$, whereas ECT between $\QDii$ and $\QDiii$ drains no such current to ground via S$_2$. The opposite sequence, marked by the blue arrow, takes place with $\mu_3 \approx -\mu_2 \approx -\mu_1$ and shows similar behavior. We emphasize that this coupling between all three sites gives rise to a non-local transport feature. For example, we observe in Fig.~\ref{fig:4}c,d that $\IL$ is strongly modulated by $\QDiii$, two sites away. 

\begin{figure}[ht!]
    \centering
    \includegraphics[width=\columnwidth]{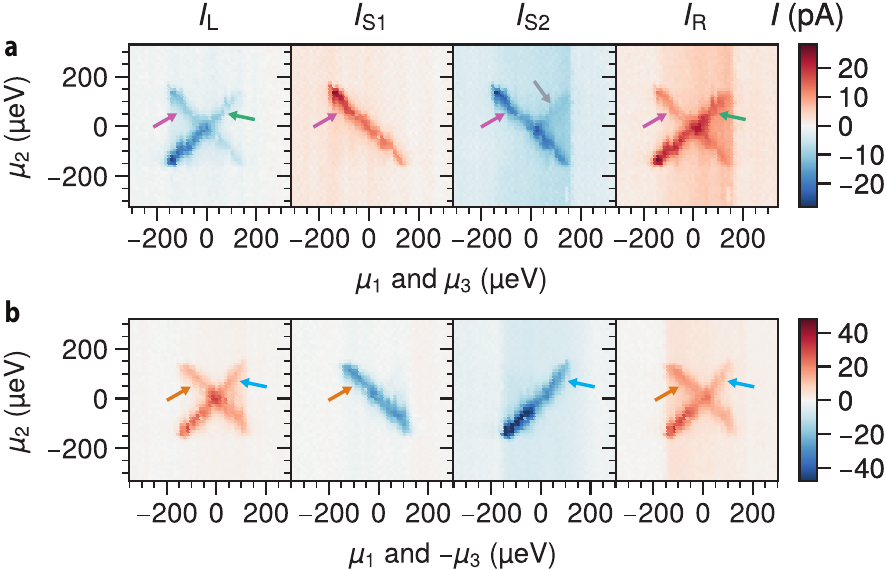}
    \caption{\textbf{a.} Current through the device as a function of $\mu_2$ and jointly $\mu_1$ and $\mu_3$, which are set to equal values, measured with antisymmetric bias configuration. \textbf{b.} Current through the device as a function of $\mu_2$ and jointly $\mu_1$ and $\mu_3$, which are set to opposite values, measured with symmetric bias configuration. Note that the color of the arrows corresponds to the color of the processes in Fig.~\ref{fig:4} (the grey arrow highlights the faint SPT process discussed in Fig.~\ref{fig:3}).}
    \label{fig:5}
\end{figure}


Sequential CAR and ECT processes across the entire device require the simultaneous tuning of the chemical potential of all three QDs. 
To better visualize sequential CAR and ECT processes, we  note that both sequential ECT processes and sequential CAR processes always require $\mu_1 = \mu_3$, whereas CAR followed by ECT and ECT followed by CAR require $\mu_1 = -\mu_3$. In Fig.~\ref{fig:5}, we measure the currents through the devices while tuning the QDs to follow these constraints. Fig.~\ref{fig:5}a was measured by setting $\VR = -\VL = \SI{150}{\micro V}$.  $\VQDi$ and $\VQDiii$ were swept together, imposing $\mu_1 = \mu_3$ for the full measurement (see Fig.~S4 in Supplementary Information for details regarding the tuning of chemical potential). Fig.~\ref{fig:5}a features two diagonal lines. 
The positive-slope diagonal, compatible with $\mu_1 = \mu_2 = \mu_3$, is prominent in panels $\IL$ and $\IR$ only, allowing us to attribute it to sequential ECT processes. The negative-slope diagonal, compatible with $\mu_1 = -\mu_2 = \mu_3$, appears in all panels and is associated with sequential CAR processes. The measurements in Fig.~\ref{fig:5}b were conducted with $\VR = \VL = \SI{150}{\micro V}$. Here, $\VQDi$ and $\VQDiii$ are varied together, while imposing $\mu_1 = -\mu_3$. Similarly to the previous scenario, measured currents feature a positive-slope diagonal alongside a negative-slope one. 
Here, the positive-slope diagonal involves $\ISii$ as expected for ECT followed by CAR. 
The negative-slope diagonal involves $\ISi$ instead, as required by CAR followed by ECT. 
In summary, the results of Fig.~\ref{fig:4}~and~\ref{fig:5} show how all four possible compositions of CAR and ECT mediate transport through the entire Kitaev chain device. 

\section*{Conclusion}
In summary, we have fabricated and measured an InSb-Al device in a three-site Kitaev geometry that includes QDs separated by semiconductor-superconductor hybrids. We have shown the signatures of CAR and ECT between the two pairs of neighboring QDs exhibiting the expected charge and energy conservation for each respective process. We have further demonstrated control over sequential CAR and ECT processes involving all QDs by tuning the biases applied to the normal leads and the chemical potential of the QDs. Our measurements demonstrate the possibility of extending the known Kitaev chain physics to longer multi-site chains, leaving the fine-tuning of interdot couplings at finite magnetic field to future work. Finally, we note that this experiment demonstrates a general platform enabling long-range entanglement in condensed matter systems~\cite{Choi.2000,Recher.2001}. For instance, we note that the extension to three sequential CAR events involving four QDs realizes a simple entanglement swapping scheme~\cite{bennett1993teleporting, zukowski1993event}.



\section*{Acknowledgements}
This work has been supported by the Dutch Organization for Scientific Research (NWO) and Microsoft Corporation Station Q. 
We wish to acknowledge Srijit Goswami, Francesco Zatelli and Greg Mazur for useful discussions and Ghada Badawy, Sasa Gazibegovic and Erik~P.~A.~M. Bakkers for the nanowire growth.

\section*{Author contributions}

AB, XL, JCW, and DvD fabricated the devices. AB and XL performed the electrical measurements with help from QW and SLDtH. AB and TD designed the experiment and analyzed the data. TD and LPK supervised the project. AB, TD, and LPK prepared the manuscript with input from all authors.

\section*{Data availability}
All raw data in the publication and the analysis code used to generate figures are available at 
\url{https://doi.org/10.5281/zenodo.8021184}. In the same repository, we share in addition a complete dataset of three-dimensional current measurements as a function of $\VQDi, \VQDii$ and $\VQDiii$ for all combinations of symmetric and anti-symmetric biases ($\Vb = \pm \VL = \pm \VR$). We share a similar dataset for a second device as well (see also Fig.~S5 in Supplementary Information). We include GIF animations for better visualization.

%


\end{document}


\title{Supplementary Information --- Crossed Andreev reflection and elastic co-tunneling in a three-site Kitaev chain nanowire device}

\author{Alberto~Bordin}
\author{Xiang~Li}
\author{David~van~Driel}
\author{Jan~Cornelis~Wolff}
\author{Qingzhen~Wang}
\author{Sebastiaan~L.~D.~ten~Haaf}
\author{Guanzhong~Wang}
\author{Nick~van~Loo}
\author{Leo~P.~Kouwenhoven}
\email{l.p.kouwenhoven@tudelft.nl}
\author{Tom~Dvir}
\email{tom.dvir@gmail.com}
\affiliation{QuTech and Kavli Institute of NanoScience, Delft University of Technology, 2600 GA Delft, The Netherlands}

\date{\today}

\maketitle

\renewcommand\thefigure{S\arabic{figure}}
\setcounter{figure}{0}

\section*{Nanofabrication details}

The device is fabricated by depositing an InSb nanowire with a micromanipulator on top of a keyboard of 11 pre-patterned Ti/Pd gates. Nanowire and gates are separated by a double-layer dielectric deposited with ALD: \SI{10}{nm} of Al$_2$O$_3$ followed by \SI{10}{nm} of HfO$_2$. Two superconducting Al contacts $\SL$ and $\SR$ are deposited with the shadow-wall lithography technique~\cite{Heedt.2021}. The Al is deposited at a temperature of 140K and a rate of \SI{0.05}{\angstrom/s}, alternating the deposition angle between 45$^\circ$ and 15$^\circ$ with respect to the substrate. This produces a uniform Al coating, \SI{9}{nm} thick, on three over six facets of the nanowire, which has a hexagonal cross-section~\cite{Badawy.2019}. Without breaking the vacuum, the Al is covered by \SI{7}{nm} of Al$_2$O$_3$. Finally, two normal Cr/Au contacts $\NL$ and $\NR$ are deposited at the two edges of the nanowire. Prior to Al deposition, the native nanowire oxide is removed with a gentle H cleaning~\cite{Heedt.2021}; prior to Cr/Au deposition, the oxide is removed with Ar milling.

\section*{Setup discussion}

As pointed out in the main text, every contact of the four-terminal device is connected to an independent voltage source ($\VL$, $\VSi$, $\VSii$, $\VR$) and current meter ($\IL$, $\ISi$, $\ISii$, $\IR$). 
Due to Kirchoff's law, a minimal setup requires only 3 voltage differences and 3 current meters. We choose to set up 4 for symmetry reasons. Moreover, such a redundant setup allows checking that the sum of all currents is compatible with the noise floor, ruling out potential leakage currents to the gates.

Through the main text, the voltage biases on the normal leads are set to $\pm \Vb=\pm \SI{150}{\micro V}$. This value is smaller than the energy of the lowest ABS and large enough to have an appreciable bias window $\abs{e\Vb}$ for the QD chemical potentials.

\subsection*{Finger gate settings}

\begin{figure}[h]
    \centering
    \includegraphics[width=0.65\textwidth]{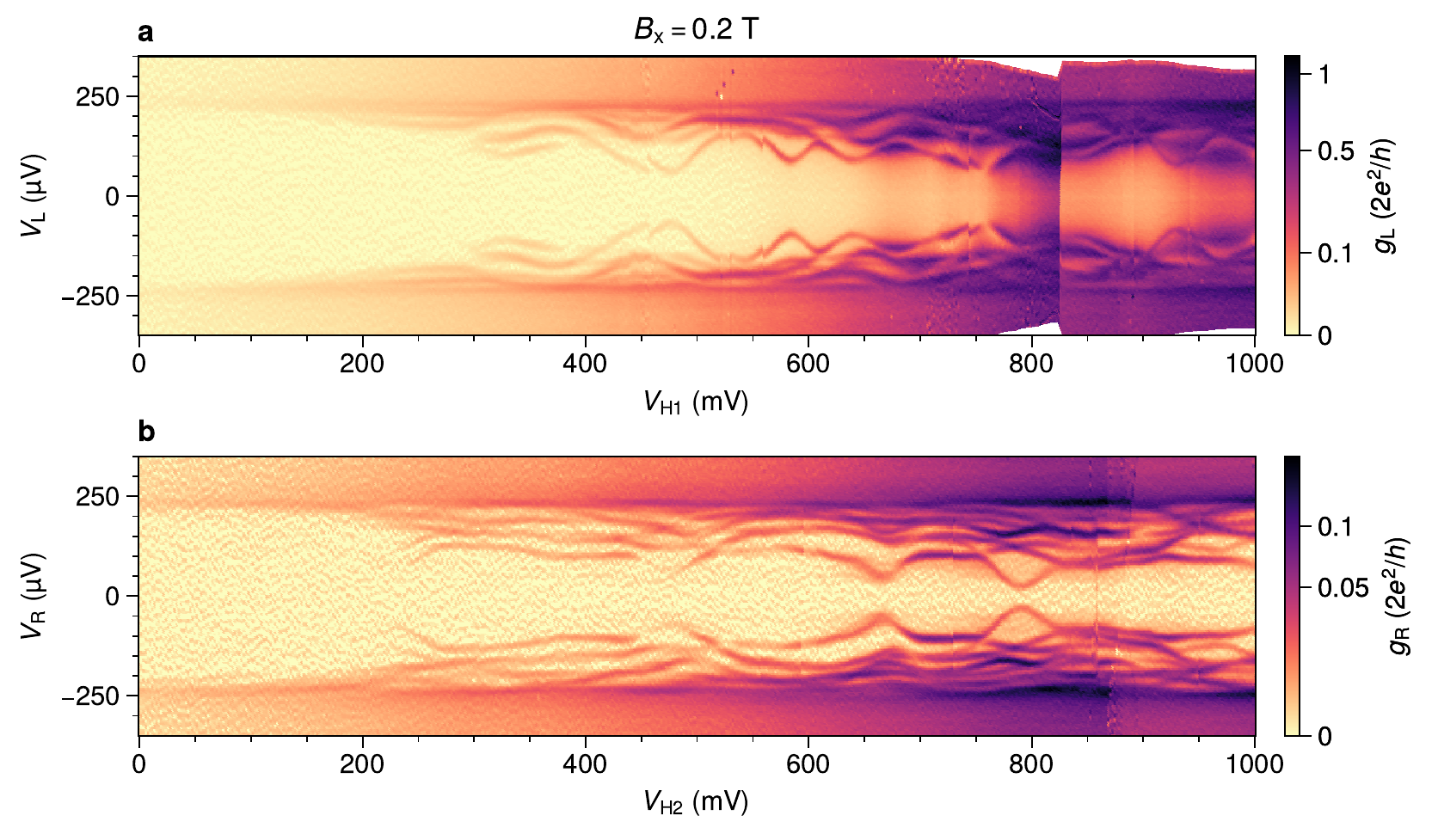}
    \caption{\textbf{a.} Tunneling spectroscopy of the first hybrid as a function of $V_\mathrm{H1}$. \textbf{b.} Tunneling spectroscopy of the second hybrid as a function of $V_\mathrm{H2}$. For both panels, $B_\mathrm{x} = \SI{0.2}{T}$, $V_\mathrm{1L} = \VQDi = \SI{0.5}{V}$, $V_\mathrm{1R} = \SI{-0.15}{V}$, $V_\mathrm{2L} = \VQDii = V_\mathrm{2R} = 0$, $V_\mathrm{3L} \approx \SI{-0.1}{V}$, $\VQDiii = V_\mathrm{3R} = \SI{0.5}{V}$. $g_\mathrm{L} \equiv \frac{dI_\mathrm{L}}{dV_\mathrm{L}}$ and $g_\mathrm{R} \equiv \frac{dI_\mathrm{R}}{dV_\mathrm{R}}$ are calculated by taking the numerical derivative after applying a Savitzky-Golay filter of window length 3 and polynomial order 1. $\VL, \VR, g_\mathrm{L}, g_\mathrm{R}$ take into account a series resistance \Rs due to the dilution refrigerator lines and measurement electronics.}.
    \label{supp:1}
\end{figure}

We denote the voltages applied to the 11 finger gates with $V_\mathrm{1L}$, $\VQDi$, $V_\mathrm{1R}$, $V_\mathrm{H1}$, $V_\mathrm{2L}$, $\VQDii$, $V_\mathrm{2R}$, $V_\mathrm{H2}$, $V_\mathrm{3L}$, $\VQDiii$, $V_\mathrm{3R}$ from left to right. We set the hybrid gates $V_\mathrm{H1}$ and $V_\mathrm{H2}$ such that both hybrids hold Andreev bound states (ABSs). Fig.~\ref{supp:1} shows bias spectroscopy of the two hybrids at $B_\mathrm{x} = \SI{0.2}{T}$ ($x$ is the direction along the length of the nanowire), where ABSs are easier to see due to their high $g$-factor $\sim 20$~\cite{Wang.2022}. ABSs appear for $V_\mathrm{H1} > \SI{0.3}{V}$ and for $V_\mathrm{H2} > \SI{0.2}{V}$.

\begin{figure}[h]
    \centering
    \includegraphics[width=0.5\textwidth]{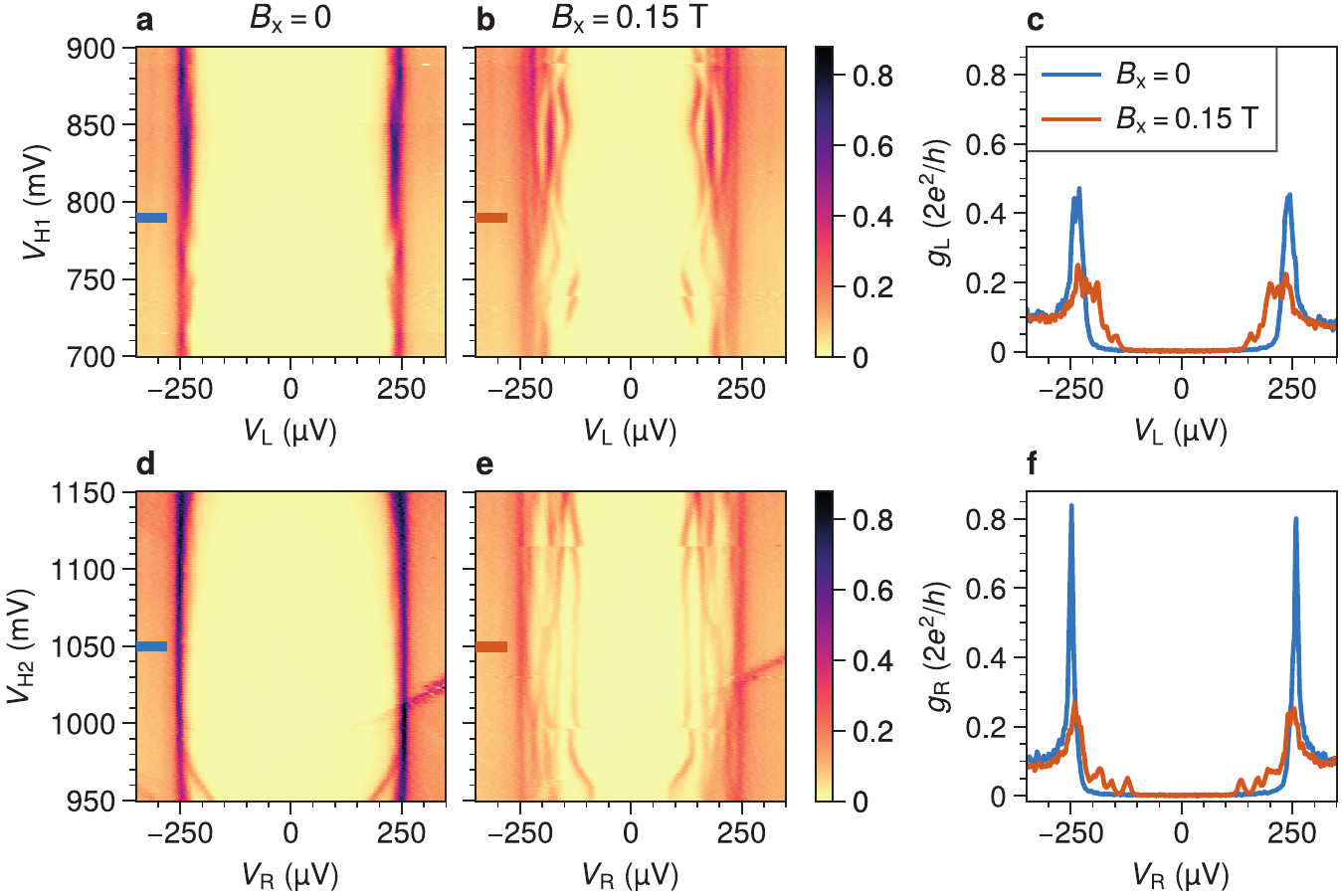}
    \caption{\textbf{a., b., c.} Tunneling spectroscopy of the first hybrid. We fix $V_\mathrm{H1} = \SI{0.79}{V}$ for all other figures (with small changes within $\SI{0.05}{V}$, see the linked repository for further details). \textbf{d., e., f.} Tunneling spectroscopy of the second hybrid. $V_\mathrm{H2} = \SI{1.05}{V}$ for all other figures. In all panels, $V_\mathrm{1L} = \SI{0.5}{V}$, $\VQDi = \SI{0.4}{V}$, $V_\mathrm{1R} = \SI{-0.14}{V}$, $V_\mathrm{2L} = \SI{-0.02}{V}$, $\VQDii = \SI{0.472}{V}$, $V_\mathrm{2R} =\SI{-0.015}{V}$, $V_\mathrm{3L} = \SI{-0.055}{V}$, $\VQDiii = \SI{0.49}{V}$, $V_\mathrm{3R} = \SI{0.5}{V}$. $g_\mathrm{L}$ and $g_\mathrm{R}$ are calculated by taking the numerical derivative after applying a Savitzky-Golay filter of window length 3 and polynomial order 1. $\VL, \VR, g_\mathrm{L}, g_\mathrm{R}$ take into account a series resistance \Rs. Panels \textbf{c.} and \textbf{f.} report the same data of Fig.~\ref{fig:1}c,d.}
    \label{supp:2}
\end{figure}

\newpage

\begin{figure}[h]
    \centering
    \includegraphics[width=0.75\textwidth]{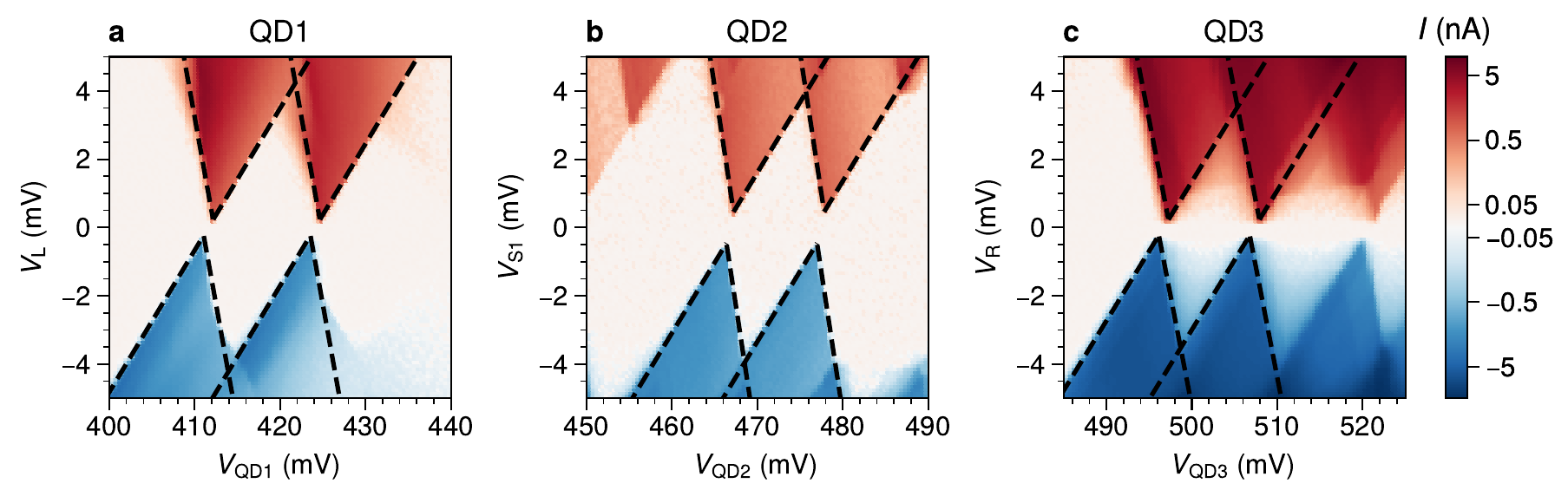}
    \caption{\textbf{a., b., c.} Coulomb diamonds for QDs 1-3. It is the same data reported in the main text in Fig.~1e-g but with a logarithmic colorbar scale in order to visualize small currents inside the Coulomb blockaded regions. A linear interpolation between $-0.03$ and $+0.03$ avoids the logarithm divergence for small values. All Coulomb diamonds are very sharp, a signature of high tunneling barriers. Only $\QDiii$, the QD where we measured the strongest local Andreev reflection current, shows some current leaking inside the Coulomb blockade diamond. Tunneling gates are set to $V_\mathrm{1L} = \SI{-0.123}{V}$, $V_\mathrm{1R} = \SI{-0.14}{V}$, $V_\mathrm{2L} = \SI{-0.02}{V}$, $V_\mathrm{2R} =\SI{-0.015}{V}$, $V_\mathrm{3L} = \SI{-0.053}{V}$, $V_\mathrm{3R} = \SI{-0.395}{V}$.}
    \label{supp:3}
\end{figure}

Since ECT followed by local Andreev reflection can mimic CAR~\cite{Schindele.2014}, we set the gates in order to minimize local Andreev reflection while keeping a detectable CAR signal. In order to do so, the hybrid gates are fine-tuned to the values of Fig.~\ref{supp:2}, where both ECT and CAR signals are strong (see Fig.~2 in the main text), while the tunnelling barriers defining the quantum dots are kept as high as possible in order to suppress local Andreev reflection. Typical barrier gate voltages are reported in Fig.~\ref{supp:3}. Numbers might vary a little from Fig. to Fig. (within $\SI{0.01}{V}$), all values are available in the linked repository.

\pagebreak
\newpage
\subsection*{$\mu_1=\mu_3$ and $\mu_1=-\mu_3$ tuning}

\begin{figure}[h]
    \centering
    \includegraphics[width=0.5\textwidth]{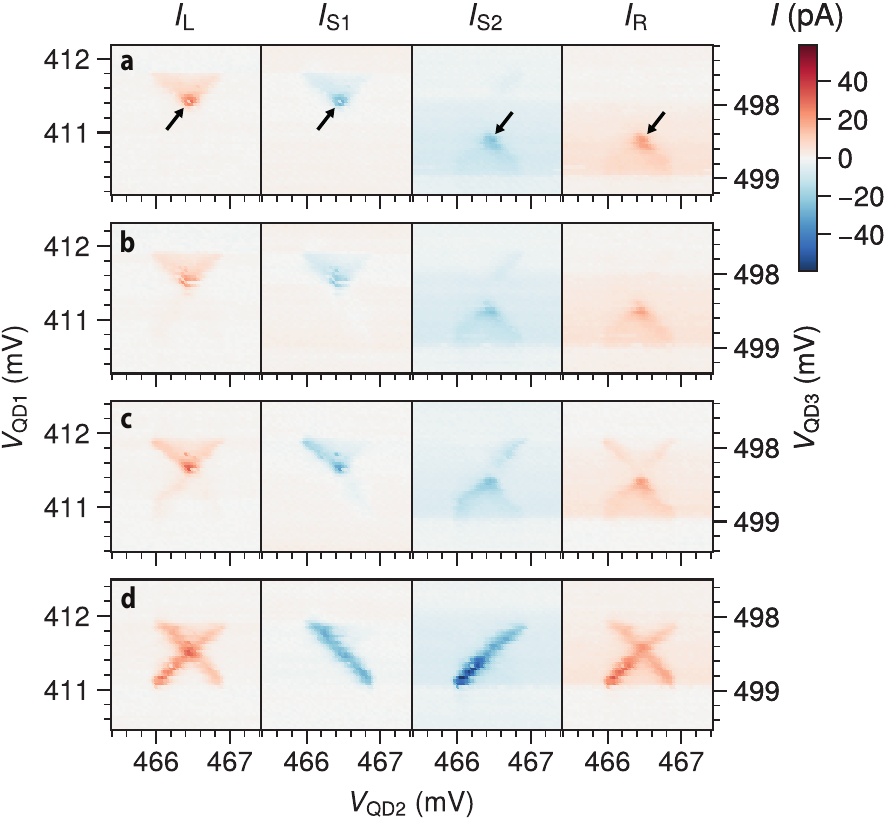}
    \caption{$\mu_1=-\mu_3$ tuning example. \textbf{a., b., c., d.} Sequence of current measurements as a function of $\VQDi$ and $\VQDii$ with $\VQDi = -\VQDiii + V_+$. $V_+$ is varied in steps of $\SI{0.2}{mV}$ from $V_+=\SI{909.4}{mV}$ (panel a) to $V_+=\SI{910.0}{mV}$ (panel d). Black arrows highlight the resonant CAR and ECT processes happening on the left when $\mu_1=\mu_2=0$ (panels $\IL$ and $\ISi$) and on the right when $\mu_2=\mu_3=0$ (panels $\ISii$ and $\IR$). As $V_+$ increases, the left and right resonant CAR and ECT current spots get closer and closer until they align in panel d, where $\mu_1=\mu_2=\mu_3=0$ at the center of the cross.}
    \label{supp:4}
\end{figure}

In Fig.~5 of the main text, the chemical potentials of $\QDi$ and $\QDiii$ are set to be either equal (Fig.~5a) or opposite (Fig.~5b). Both situations require careful calibration. Fig.~\ref{supp:4} shows how $\VQDi$ and $\VQDiii$ can be finely tuned in order to set $\mu_1 = -\mu_3$ (the $\mu_1 = \mu_3$ case is analogous).
The chemical potentials are related to the gate voltages via $\mu_1 = \alpha_1 (\VQDi - V_{1})$ and $\mu_3 = \alpha_3 (\VQDiii - V_{3})$, where $\alpha_1=0.32$ and $\alpha_3=0.31$ are the lever arms and $V_{1}\approx\SI{412}{mV}$ and $V_{3}\approx\SI{497}{mV}$ are offsets. These values are extracted from the Coulomb diamonds of Fig.~1.
Assuming equal lever arms, $\mu_1=-\mu_3$ if $(\VQDi - V_{1}) = -(\VQDiii - V_{3})$, which means $\VQDi + \VQDiii = \textrm{constant} \equiv V_+$. Fig.~\ref{supp:4} shows the measured currents as the value of $V_+$ is varied in small steps until the condition $\mu_1=-\mu_3$ is reached.

\pagebreak
\newpage
\section*{Extended data}

\begin{figure}[h]
    \centering
    \includegraphics[width=0.75\textwidth]{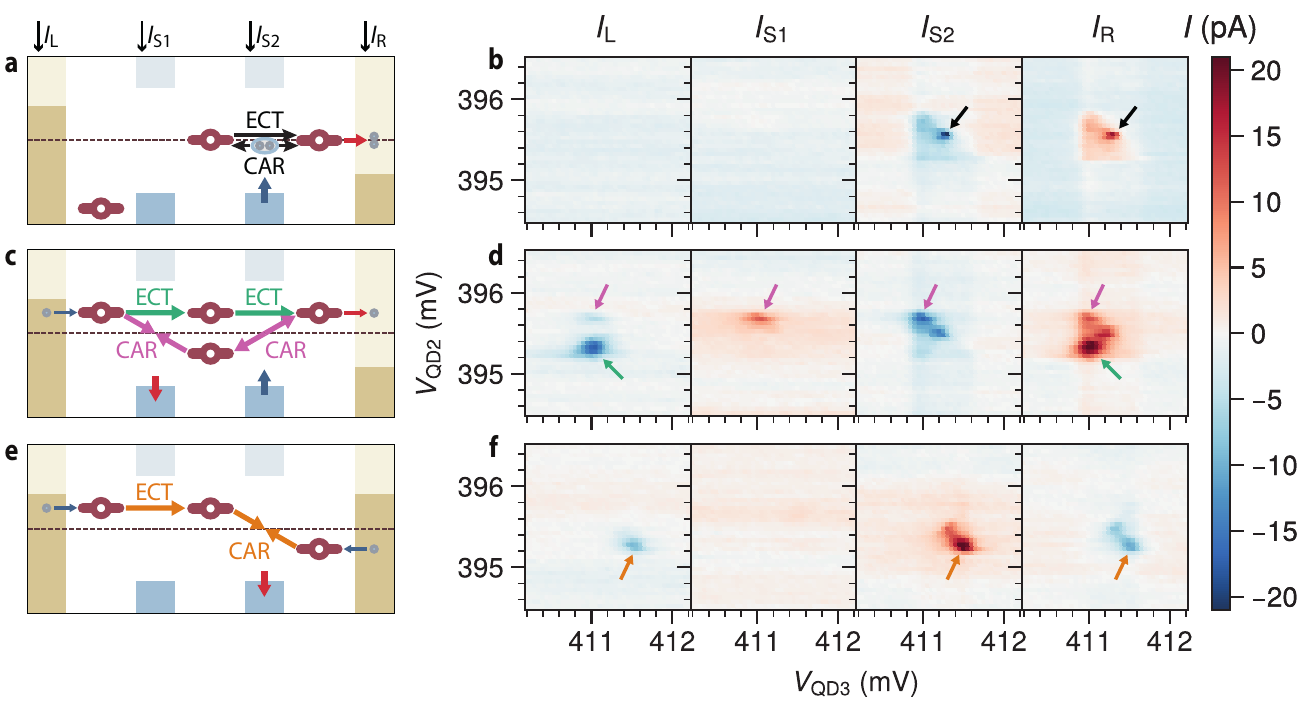}
    \caption{Signatures of sequential CAR and ECT processes in a second device. \textbf{a.} Schematic illustration of resonant CAR and ECT tunneling. \textbf{b.} Current through the device as a function of $\VQDii$ and $\VQDiii$, with $\mu_1$ out of the bias window ($\VQDi = \SI{294.65}{mV}$). \textbf{c.} Schematic illustration of sequential ECT and sequential CAR processes. \textbf{d.} Current through the device as a function of $\VQDii$ and $\VQDiii$, with $\mu_1$ within the bias window ($\VQDi=\SI{295.15}{mV}$), measured with an antisymmetric bias configuration \textbf{e.} Schematic illustration of ECT followed by CAR. \textbf{f.} Current through the device as a function of $\VQDii$ and $\VQDiii$, with $\mu_1$ within the bias window ($\VQDi=\SI{295.15}{mV}$), measured with a symmetric bias configuration. The arrow colors are chosen as in the main text. In panels b, d, f, a constant background current is subtracted in every $I$ panel for better visibility, see the code in the linked repository for further details.}
    \label{supp:5}
\end{figure}

In Fig.~\ref{supp:5} we report signatures of sequential CAR and ECT measured in a second device. Additional data is shared in the linked repository, including three-dimensional current measurements as a function of $\VQDi, \VQDii$ and $\VQDiii$ with $\VL$ and $\VR$ set in all possible symmetric and anti-symmetric configurations (with $\Vb = \SI{100}{\micro V}$). We include GIF images for data visualization.

